# Electrical conductance of charged nanopores

Yoav Green*

Department of Mechanical Engineering, Ben-Gurion University of the Negev, Beer-Sheva 8410501, Israel

A nanopore's response to an electrical potential drop is characterized by its electrical conductance, $\tilde{G}$. It has long been thought that at low concentrations, the conductance is independent of the electrolyte concentration, $\tilde{c}_0$, such that $\tilde{G} \sim \tilde{c}_0^0$. It has been recently demonstrated that surface charge regulation changes the dependency to be $\tilde{G} \sim \tilde{c}_0^\alpha$ whereby the slope typically takes the values $\alpha = \frac{1}{3}$ or $\frac{1}{2}$. Yet, experiments have observed slopes of $\frac{2}{3}$ and 1 suggesting that additional mechanisms, such as convection and slip-lengths, appear. We show that the inclusion of convection doesn't vary the slope, while the inclusion of a slip length doubles the slope value. Here, we elucidate the interplay between surface charge regulation, convection, and slip lengths. We show that when all effects are accounted for $\alpha$ can take any value between 0 and 1. This result is of utmost importance in designing any electro-kinetically driven nanofluidic system characterized by its conductance.

* yoavgreen@bgu.ac.il   ORCID number: 0000-0002-0809-6575

**Introduction.** The discovery of new materials and the development of more advanced fabrication methods results in system sizes that are ever decreasing [1]. With this comes the potential to enhance our understanding of nanoscale physics and, in parallel, revolutionize current technological setups. Of particular interest is the transport of ions across these nanoscale systems that are found in desalination [2–7] and energy harvesting [8–17] systems, as well as biosensing [18–22], fluid-based electrical diodes [23–31], and various physiological phenomena [32–35]. However, numerous challenges related to scalability, fabrication technology, and elucidation of the unknown fundamental physics at these small scales [7–9,36–38] remain to be overcome.

It is known that a plethora of mechanisms, unique to nanoscale systems, determine the system's overall response, yet the interplay of these mechanisms is not understood. Specifically, this work addresses the interplay of surface charge regulation, bulk convection, and slip-length induced convection on the nanopore conductance. Notably, we will show that a combination of all three effects allows the slope of the conductance to surpass the maximal slope of $\frac{1}{2}$ predicted by surface charge regulation. We will show that once all three phenomena are accounted for, the slope can take any value between 0 and 1. Notably, the model suggested in this work is free of any fitting parameters such that the response is determined solely by the various system parameters.

**Ohmic Conductance.** The electrical conductance, $\tilde{G}$, is the ratio of the electrical current, $\tilde{I}$, to the electrical potential drop, $\tilde{V}$ (i.e. $\tilde{G} = \tilde{I}/\tilde{V}$). Stein et al.'s [39] pioneering work showed that the Ohmic conductance of nanochannels and nanopores [**Figure 1**(a)] behaved in a peculiar manner. At high bulk concentrations, $\tilde{c}_0$, when the electric doubles layers (EDLs, defined below) do not overlap, the conductance increases linearly with the bulk concentration ($\tilde{G}_{high} \sim \tilde{c}_0$). At low concentrations, when the EDLs overlap, the conductance saturates to a constant value that depends on the surface charge density, $\tilde{\sigma}_s$, but is independent of the concentration ($\tilde{G}_{low} \sim \tilde{c}_0^0 \sim \tilde{\sigma}_s$). The red line with squares in **Figure 2** depicts the behavior schematically and is given by the well-known equation [40,41]

$$\tilde{G}_{Ohmic} = \tilde{\kappa}_{cond}\sqrt{4 + \left(\frac{\tilde{N}}{\tilde{c}_0}\right)^2}\frac{\pi\tilde{a}^2}{\tilde{L}}, \qquad (1)$$

where $\tilde{\kappa}_{cond} = z^2\tilde{F}^2\tilde{D}\tilde{c}_0/(\tilde{R}_g\tilde{T})$ is the conductivity, $\tilde{R}_g$ is the universal gas constant, $\tilde{T}$ is the temperature, $\tilde{F}$ is the Faraday constant, $\tilde{D}$ is the diffusion coefficient, and $z$ is the valence. The pore radius and pore length are given by $\tilde{a}$ and $\tilde{L}$, respectively [**Figure 1**(a)]. Here $\tilde{N} = -(2\tilde{\sigma}_s)/(\tilde{a}\tilde{F}z)$ represents the average excess counterion concentration due to the surface charge density. Equation (1) holds for a symmetric electrolyte where the counterion and coions have equal diffusion coefficients ($\tilde{D}_\pm = \tilde{D}$) and opposite valences ($z_\pm = \pm z$) and a channel with a large aspect ratio $\tilde{L}/\tilde{a} \gg 1$. Also, the effects of convection are assumed negligible. In this work, similar to Eq. (1), we will consider a pore with a large aspect ratio ($\tilde{L}/\tilde{a} \gg 1$). However, in contrast to Eq. (1), we will also account for the effects of convection.

At high concentrations, $\tilde{N} \ll \tilde{c}_0$, the conductance is linear with the concentration, $\tilde{G}_{Ohmic,high} \sim \tilde{\kappa}_{cond} \sim \tilde{c}_0$ (**Figure 2**). In contrast, at low concentrations, $\tilde{N} \gg \tilde{c}_0$, one finds that the conductance

$$\tilde{G}_{Ohmic,low} = \frac{2\pi\tilde{a}}{\tilde{L}}\frac{z\tilde{F}\tilde{D}}{\tilde{R}_g\tilde{T}}|\tilde{\sigma}_s|, \qquad (2)$$

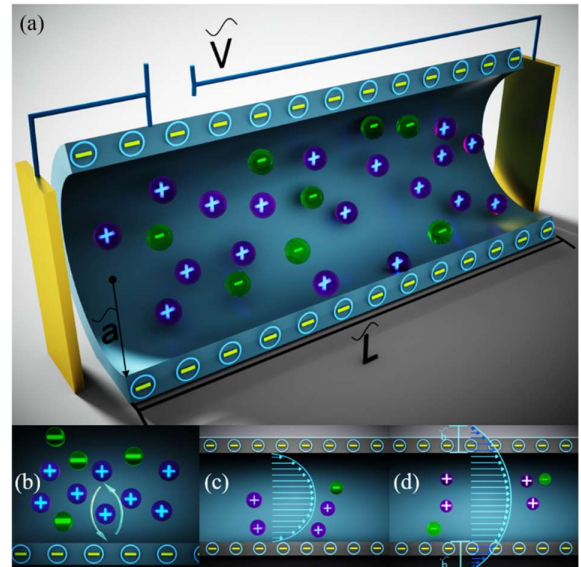

**Figure 1**. (a) Schematic representation of a negatively charged long nanotube ($\tilde{L} \gg \tilde{a}$) under an applied voltage, $\tilde{V}$. Due to the negative surface charge density, $\tilde{\sigma}_s$, there is an excess of positive counterions, represented by purple spheres, over the negative coions, represented by green spheres. This work focuses on the case of a highly selective channel ($\varepsilon = \tilde{\lambda}_D/\tilde{a} \gg 1$) which corresponds to the case of few negative ions. (b) The surface charge density is regulated by $H_+$ (not drawn here). A schematic profile of (c) a no-slip velocity profile and (d) a velocity profile with a slip length $\tilde{b}$.

is explicitly concentration-independent such that $\alpha = 0$ (**Figure 2**). However, recent works have suggested that the surface charge is concentration-dependent through a mechanism known as surface charge regulation [43–46]. Surface charge regulation couples the surface charge density to the bulk electrolyte such that $\tilde{\sigma}_{s,\alpha} \sim \tilde{c}_0^{\alpha}$. Here $\alpha$ is the exponent of the power-law that determines the slope of the conductance. Surface charge regulation predicts three distinct solutions: 1) $\tilde{\sigma}_{s,0} = -\tilde{F}\tilde{n}/\tilde{N}_a$; 2) $\tilde{\sigma}_{s,\frac{1}{3}} = -(2\tilde{\varepsilon}_0 \varepsilon_r \tilde{R}_g \tilde{T} z \tilde{c}_0 \gamma/\beta)^{1/3}$ [43]; 3) $\tilde{\sigma}_{s,\frac{1}{2}} = -(\frac{1}{2}\tilde{a}\tilde{F}\tilde{c}_0 z\gamma/\beta)^{1/2}$ [47]; where $\tilde{\varepsilon}_0$ and $\varepsilon_r$ are the permittivity of free space and the relative permittivity, and $\gamma$ and $\beta$ are defined below Eq. (4). Insertion of these three expressions into Eq. (1) leads to the three curves shown in **Figure 2** with the appropriate slopes of $\alpha = 0, \frac{1}{3}$ and $\frac{1}{2}$. All three scenarios show remarkable correspondence to direct 2D numerical simulations [42].

**Surface charge regulation**. Smeets et al. [48] demonstrated that under certain conditions, the conductance exhibited a non-zero slope. This slope was attributed to a concentration dependency of the surface charge density. This mechanism has started to receive increased interest in recent years where it has been suggested that the surface charge density is regulated [**Figure 1**(b)] through the Langmuir isotherm [43–46,49]

$$\tilde{\sigma}_s = -\frac{\tilde{F}\tilde{n}}{\tilde{N}_a}\left[1 + 10^{\text{pK}-\text{pH}_\infty}\exp\left(-\frac{\tilde{\varphi}_s}{\tilde{\varphi}_{th}}\right)\right]^{-1}. \quad (3)$$

Here, $\tilde{N}_a$ is Avogadro's constant, $\tilde{n}$ is the maximal number of ionizable sites per unit area, $p$K is the disassociation constant, $p$H$_\infty$ is the $p$H in the bulk concentration, and $\tilde{\varphi}_{th} = \tilde{R}_g\tilde{T}/\tilde{F}z$ is the thermal potential. Notably, the surface charge density is related to the electric potential at the surface $\tilde{\varphi}_s$. At low concentrations, the surface potential is $\tilde{\varphi}_s = -\tilde{\varphi}_{th}\ln[\varepsilon^2\sigma_s(\sigma_s - 4)]$ [42], where $\varepsilon = \tilde{\lambda}_D/\tilde{a} = [(\tilde{\varepsilon}_0\varepsilon_r\tilde{R}_g\tilde{T})/(2\tilde{F}^2z^2\tilde{c}_0\tilde{a}^2)]^{1/2}$ is the normalized Debye length [or electric double layer (EDL)], and the surface charge density has been normalized by a characteristic value $\tilde{\sigma}_d = \tilde{\varepsilon}_0\varepsilon_r\tilde{\varphi}_{th}/\tilde{a}$ such that $\sigma_s = \tilde{\sigma}_s/\tilde{\sigma}_d$. In this work, dimensional quantities are denoted with tildes, while non-dimensional quantities are without tildes.

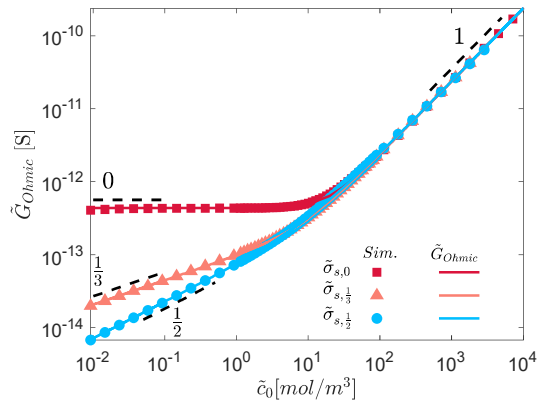

**Figure 2**. (a) Ohmic conductance versus concentration [Eq. (1)] for the three surface charge densities: $\tilde{\sigma}_{s,0}, \tilde{\sigma}_{s,\frac{1}{3}}, \tilde{\sigma}_{s,\frac{1}{2}}$. Theory is denoted by lines, and simulations are denoted by markers. Details on the simulations and the values for the simulations are given in Ref. [42].

Inserting, $\tilde{\varphi}_s$ into Eq. (3) yields a third-order polynomial that determines the non-dimensional surface charge density

$$\sigma_s^3 - 4\sigma_s^2 + (\beta\varepsilon^2)^{-1}\sigma_s + (\beta\varepsilon^2)^{-1}\gamma = 0. \quad (4)$$

Here $\beta = 10^{\text{pK}-\text{pH}_\infty}$ and $\gamma = \tilde{F}\tilde{n}/(\tilde{N}_a\tilde{\sigma}_d)$. The general solution for Eq. (4) is not tractable, yet three solutions are immediate. When $\beta\varepsilon^2 \ll 1$, the surface charge is concentration-independent $\sigma_{s,0} = -\gamma$. When $\beta\varepsilon^2 \gg 1$, two solutions concentration solutions are recovered [42]: $\sigma_{s,\frac{1}{3}} = -[\gamma/(\beta\varepsilon^2)]^{1/3}$ and $\sigma_{s,\frac{1}{2}} = -\frac{1}{2}[\gamma/(\beta\varepsilon^2)]^{1/2}$. Upon dimensionalization, these three terms recapitulate the three terms given below Eq. (2). Since both $\tilde{\sigma}_{s,\frac{1}{3}}$ [43] and $\tilde{\sigma}_{s,\frac{1}{2}}$ [44] have already been compared with experiments and shown to have excellent correspondence, we do not conduct such a comparison. Instead, here we have focused on showing that these three different models are derived from the same equation. The form of Eq. (4) suggests that depending on the parameters $\gamma$ and $\beta\varepsilon^2$, the surface charge varies continuously from one case to the other. In fact, Uematsu et al. [45] demonstrated, via numerical simulations, that the slope transitions continuously between 0 to $\frac{1}{2}$.

Uematsu et al. [45] numerically solved the 1D Poisson-Nernst-Plank equations along with the Langmuir isotherm [Eq. (3)]. They investigated how the slope of the conductance varied with the (pK, pH$_\infty$, $\tilde{c}_0, \tilde{n}, \tilde{a}$) phase space. In particular, in their simulations, they set $\tilde{a}, \tilde{n}, p$K and investigated the effects of pH$_\infty$ and $\tilde{c}_0$. They numerically calculated the conductance for each configuration in their $\tilde{c}_0 - $pH phase space using $\alpha = d(\ln\tilde{G})/d(\ln\tilde{c}_0)$. In their numerical simulations, they considered both low and high concentrations. Unexpectedly at high concentrations, they showed that the slope was 1 (as shown in **Figure 2**), while at low concentrations, the slope $\alpha$ varied from 0 to $\frac{1}{2}$. **Figure 3** is our recapitulation of the low-concentration results of Figure 1 in Uematsu et al. [45]. We use the Newton-Raphson method to numerically evaluate Eq. (4) for $\tilde{\sigma}_s$. We then insert $\tilde{\sigma}_s$ into the expression for the conductance [Eq. (2)] and calculate the slope $\alpha$. The benefits of our approach are two-fold. On the technical side, even though 1D finite-elements simulations are no longer computationally costly, scanning a 5D (or 2D) phase space can be quite burdensome. In contrast, our approach allows us to scan the phase space to *any desired resolution* in an almost *instantaneous manner*. For example, while Figure 1 of Ref. [45] is pixilated, **Figure 3** is smooth. From the physical insight perspective, our approach can also rationalize Uematsu et al.'s [45] baffling observation that lines of constant slopes (and constant colors) appeared to be given by stripes. Our theoretical approach provides a remarkable and intuitive explanation – for a given value of $\gamma$, these are lines of constant $\beta\varepsilon^2$ [42]. **Figure 3** demonstrates two key results: 1) the slope varies continuously between 0 to $\frac{1}{2}$; 2) the slopes are lines of constant $\beta\varepsilon^2$.

**Conductance with no-slip convection**. Equation (1) holds for both high and low concentrations but no longer holds upon the inclusion of convective effects. This work focuses on the low concentration-response, which exhibits slope variability, and thus we do not attempt to provide a result that holds for all concentrations. Rather, the expressions below hold only for low concentrations.

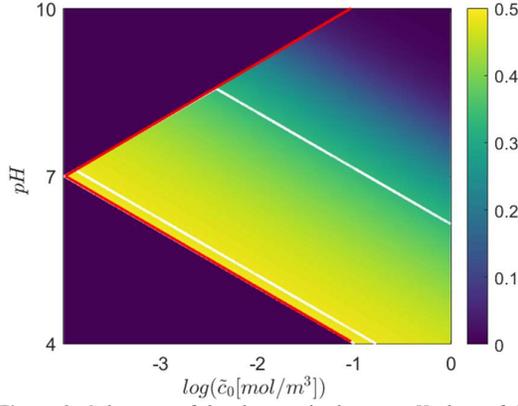

**Figure 3.** Color map of the slope $\alpha$, in the $\tilde{c}_0 - $pH plane of the Ohmic conductance, $\tilde{G}_{Ohm,low}$ [Eq. (2)]. White diagonal lines denote the lines of constant slope of $\alpha = \frac{1}{3}$ and $\alpha = 0.483$. Here, we have used the values of Figure 1 from Ref. [45]: $\tilde{a} = 35$[nm], pK = 5, and $\tilde{n} = 0.2$[nm$^{-2}$] as well as their suggested cutoffs (thick red lines).

At low concentrations, the conductance that accounts for the Ohmic contribution and the no-slip advective [**Figure 1**(c)] is [42]

$$\tilde{G}_{total,no-slip} = \tilde{G}_{Ohm,low} + \tilde{G}_{adv,no-slip}, \quad (5)$$

where the no-slip advective conductance is

$$\tilde{G}_{adv,no-sl} = -8\tilde{\kappa}_{cond}\varepsilon^2 \text{Pe}\left[\frac{\tilde{\sigma}_s}{\tilde{\sigma}_d} + 4\ln\left(1 - \frac{1}{4}\frac{\tilde{\sigma}_s}{\tilde{\sigma}_d}\right)\right]\frac{\pi\tilde{a}^2}{\tilde{L}}. \quad (6)$$

Here Pe $= \tilde{\varepsilon}_0\varepsilon_r\tilde{\varphi}_{th}^2/(\tilde{\mu}\tilde{D})$ is the Peclet number and $\tilde{\mu}$ is the fluid's viscosity. For example, for a KCl water-based electrolyte at room temperature, one finds that the Peclet number is approximately 0.45. Note that $\tilde{\kappa}_{cond}\varepsilon^2$ is concentration-independent such that $\tilde{G}_{adv,no-slip}$, too, is also explicitly concentration-independent. For large surface-charges, $\tilde{\sigma}_s \gg \tilde{\sigma}_d$, which is the typical case for highly selective nanochannels, the logarithmic term in Eq. (6) is negligible, relative to the first term, such that

$$\tilde{G}_{total,no-slip} = -4\tilde{\kappa}_{cond}\varepsilon^2 \frac{\tilde{\sigma}_s}{\tilde{\sigma}_d}\frac{\pi\tilde{a}^2}{\tilde{L}}(1+2\text{Pe}). \quad (7)$$

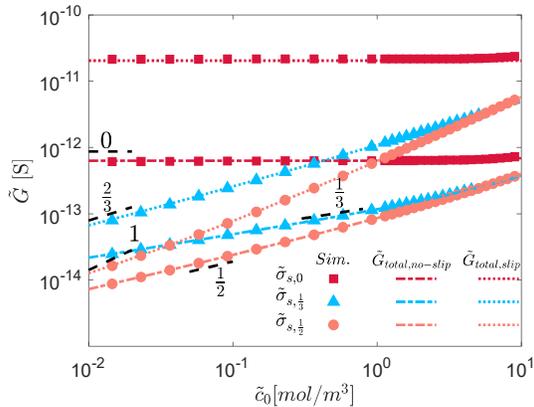

**Figure 4.** Conductance-concentrations curves for $\tilde{G}_{total,no-slip}$ [Eq. (5)] and $\tilde{G}_{total,slip}$ [Eq. (9), with a slip length is $\tilde{b} = 10\tilde{a}$]. Theory is denoted by lines, and simulations are denoted by markers. The values for the simulations are given in Ref. [42].

Observe that while the advective term results in a non-negligible increase of the conductance [50], the slope $\tilde{G}_{total,no-slip} \sim \tilde{\sigma}_s$ remains unchanged relative to the Ohmic conductance [Eq. (2)]. **Figure 4** compares Eq. (5) to numerical simulations [42] for the three cases of $\tilde{\sigma}_{s,0}$, $\tilde{\sigma}_{s,\frac{1}{3}}$, $\tilde{\sigma}_{s,\frac{1}{2}}$. The excellent correspondence confirms the prediction that no-slip convection does not change the slope.

**Conductance with slip convection.** The additional contribution of the slip length [51–54], $\tilde{b}$, [**Figure 1**(d)] to the conductance is [42]

$$\tilde{G}_{adv,slip} = 4\tilde{\kappa}_{cond}\varepsilon^2 \text{Pe}\left(\frac{\tilde{\sigma}_s}{\tilde{\sigma}_d}\right)^2 \frac{\tilde{b}}{\tilde{a}}\frac{\pi\tilde{a}^2}{\tilde{L}}. \quad (8)$$

We add Eq. (8) to Eq. (5) to get an expression for the conductance that accounts for all three contributions

$$\tilde{G}_{total,slip} = \tilde{G}_{Ohmic,low} + \tilde{G}_{adv,no-slip} + \tilde{G}_{adv,slip}. \quad (9)$$

Observe that $\tilde{G}_{adv,slip}$ further increases the conductance and that $\tilde{G}_{adv,slip}$ scales quadratically with the surface charge, $\tilde{G}_{adv,slip} \sim \tilde{\sigma}_s^2 \sim \tilde{c}_0^{2\alpha}$. Thus, if $\tilde{G}_{total,no-slip}$ has a slope $\alpha \in [0,\frac{1}{2}]$ then $\tilde{G}_{total,slip}$ has a slope $2\alpha \in [0,1]$. **Figure 4** compares the theoretical predictions of Eq. (9) to numerical simulations [42] that account for a slipe length of $\tilde{b} = 10\tilde{a}$. **Figure 4** shows that when: 1) $\tilde{\sigma}_{s,0} \sim \tilde{c}_0^0$, the slope always remains $\alpha = 0$; 2) when $\tilde{\sigma}_{s,\frac{1}{3}} \sim \tilde{c}_0^{1/3}$, the slope transitions from $\alpha = \frac{1}{3}$ to $\alpha = \frac{2}{3}$; 3) when $\tilde{\sigma}_{s,\frac{1}{2}} \sim \tilde{c}_0^{1/2}$, the slope transitions from $\alpha = \frac{1}{2}$ to $\alpha = 1$. The correspondence between simulations and theory is excellent and demonstrates the dependency of the conductance on $\tilde{\sigma}_s$ and $\tilde{b}$. Similar to $\tilde{G}_{Ohmic,low}$'s, which varies continuously from 0 to $\frac{1}{2}$ (**Figure 3**), it can be shown that $\tilde{G}_{total,slip}$ also exhibits a continuous transition from 0 to 1 that depends on the various parameters of the system (not shown here, see [42]). We note here that in a recent theoretical endeavor, Manghi et al. [49] derived a very similar (almost identical) expression to Eq. (9) [which follows from Eqs. (6)- (8)]. However, they have introduced several empirical assumptions that lack mathematical rigor and physical justification in their derivation. These issues are discussed thoroughly in Ref [42]. If one removes most of these empirical assumptions, then our Eq. (9), would be almost identical to Eq. (19) of Manghi et al. [49]. However, their analysis includes several errors [42]. Thus, the novelty of this work also lies within the analysis – here, we have predicted the doubling of the slope from $\alpha$ to $2\alpha$ for the most general scenario.

The finding that the slope doubles is of utmost experimental importance. In a recent experimental work [55], using nanotube porins, a $\frac{2}{3}$ slope was measured, suggesting that the effects of convection and slip are significant. Ref. [55] rationalized their results using the theoretical predictions of Ref. [49] who predicted a slope of $\alpha = \frac{2}{3}$. However, Ref. [49] model holds only for the specific case $\tilde{\sigma}_{s,\frac{1}{3}} \sim \tilde{c}_0^{1/3}$ ($\tilde{G}_{total,slip} \sim \tilde{c}_0^{2/3}$) whereby the universal solution, $\tilde{G}_{total,slip} \sim \tilde{c}_0^{2\alpha}$, holds for all $\alpha \in [0,\frac{1}{2}]$ (see Ref [42] for a detailed discussion between this model and that of Ref. [49]) . The doubling of the slope from $\frac{1}{2}$ to 1 could explain Ref. [3] finding that decreasing the pH from 7.5 to 3 increased the slope. This is consistent with **Figure**

**3**. Hence, the main finding of this work has been to show that the transition from a slope of $\alpha$ (for convection without slip, **Figure 3**) to $2\alpha$ (for convection with slip, now shown here, see Ref. [42]) is more robust than what has previously assumed.

**Discussion.** In recent years, the characterization of the electrical conductance of nanochannels has received increased interest. This is attributed to two experimental considerations. First, while it is slightly time-consuming, it is relatively easy to measure the conductance at various concentrations and, supposedly easy to interpret these results. Second, the change in the slope is an excellent indicator of when the channel achieves the high selectivity required for desalination and energy harvesting. Thus, the conductance provides the experimentalist with the required knowledge of the appropriate experimental conditions to consider. However, because of the numerous effects at play, interpretation of the results can be confusing. In this work, we have elucidated how the combined interplay of numerous phenomena (surface charge regulation, convection, and slip lengths) can radically change the simplest response of the nanochannel – the slope of electrical conductance as a function of the bulk concentration.

The model presented in this work is the exact solution of the fully coupled Poisson-Nernst-Planck-Stokes equation [42,56]. A minimal number of assumptions have been embedded into the model [including the assumption of a large aspect ratio ($\tilde{L} \gg \tilde{a}$)]. Direct numerical simulations of the fully coupled equations show remarkable correspondence and confirm the findings presented here and in an expanded work [42]. Namely, we have demonstrated that with surface charge regulation, the slope is not restricted to the typical three discrete values of $0$, $\frac{1}{3}$, and $\frac{1}{2}$ but rather the slope can take any value between $0$ and $\frac{1}{2}$. The inclusion of convection alone does not change the slope, but it does enhance the conductance by a factor of twice the Peclet number (which is a value that characterizes the electrolyte). The inclusion of a slip length increases the conductance and, more importantly, doubles the slope to be twice the value dictated by surface charge regulation.

The final expression $\tilde{G}_{total,slip}$ provides the interested experimentalist a vastly enhanced framework for interpreting experimental results, and a means to fit a more accurate curve with "virtually" zero fitting parameters. If the maximal number of ionizable sites per unit area, $\tilde{n}$, and the slip length, $\tilde{b}$, are known this model is entirely devoid of fitting parameters. If they are not known, they are easily fitted [42]. For example, in our numerical simulations, these values are proscribed a priori, and thus from the numerical perspective, we have zero fitting parameters.

This work serves as a stepping stone to many future works that should focus on non-trivial open questions. How does the system response change if one accounts for multiple species (that are not necessarily symmetric)? How does the response change when the assumption of a large aspect ratio ($\tilde{L} \gg \tilde{a}$) is alleviated? This last question is highly pertinent to novel 2D materials whereby $\tilde{L} \sim \tilde{a}$ or even $\tilde{L} \ll \tilde{a}$. In such a scenario, the assumption of fully-developed flows needs to be reevaluated as well as the assumption that the slip length can vary with the ratio $\tilde{a}/\tilde{L}$ [51–54]. We note that for the case of high concentrations (no EDL overlap), Yariv and Sherwood [57] showed that the system can be considered to be fully developed for $\tilde{L} \sim \tilde{a}$. However, this has yet to be shown for low-concentration (and highly selective) systems and should also be examined. This issue of fully developed profiles can also be linked to how the response of the system changes when the adjacent reservoirs are accounted. It is known that under certain conditions, the effects of access resistance and microchannel resistances are no longer non-negligible [34,58,59]. Yet, if the profiles are not fully developed, how access resistance effects are manifested requires reexamination. How does the system change if there is a breakdown of electroneutrality [60,61]? An expanded discussion of all these open questions is provided in Ref. [42].

In conclusion, here, we have delineated the interplay of surface charge regulation, convection, and slip lengths on the slope of the conductance. The results of this work are of immense importance when designing electro-kinetically based nanofluidics systems. Our model provides crucial insights for data interpretation. They also provide a means to reduce the number of time-consuming experiments and numerical simulations needed for the preliminary characterization of such systems.

**ACKNOWLEDGEMENTS**

This work was supported by the Israel Science Foundation (Grant Nos. 337/20 and 1953/20). We thank the Ilse Katz Institute for Nanoscale Science & Technology for their support.

**REFERENCES**

[1] J. P. Thiruraman, P. Masih Das, and M. Drndić, *Ions and Water Dancing through Atom-Scale Holes: A Perspective toward "Size Zero,"* ACS Nano **14**, 3736 (2020).
[2] S. K. Patel, P. M. Biesheuvel, and M. Elimelech, *Energy Consumption of Brackish Water Desalination: Identifying the Sweet Spots for Electrodialysis and Reverse Osmosis*, ACS EST Eng. (2021).
[3] R. H. Tunuguntla, R. Y. Henley, Y.-C. Yao, T. A. Pham, M. Wanunu, and A. Noy, *Enhanced Water Permeability and Tunable Ion Selectivity in Subnanometer Carbon Nanotube Porins*, Science **357**, 792 (2017).
[4] J. K. Holt, H. G. Park, Y. Wang, M. Stadermann, A. B. Artyukhin, C. P. Grigoropoulos, A. Noy, and O. Bakajin, *Fast Mass Transport Through Sub-2-Nanometer Carbon Nanotubes*, Science **312**, 1034 (2006).
[5] S. P. Surwade et al., *Water Desalination Using Nanoporous Single-Layer Graphene*, Nature Nanotechnology **10**, 459 (2015).
[6] J. Feng et al., *Single-Layer MoS$_2$ Nanopores as Nanopower Generators*, Nature **536**, 197 (2016).
[7] C. Chen and L. Hu, *Nanoscale Ion Regulation in Wood-Based Structures and Their Device Applications*, Advanced Materials **33**, 2002890 (2021).
[8] L. Bocquet, *Nanofluidics Coming of Age*, Nat. Mater. **19**, 3 (2020).
[9] N. Kavokine, R. R. Netz, and L. Bocquet, *Fluids at the Nanoscale: From Continuum to Subcontinuum Transport*, Annu. Rev. Fluid Mech. (2020).
[10] A. Siria, P. Poncharal, A.-L. Biance, R. Fulcrand, X. Blase, S. T. Purcell, and L. Bocquet, *Giant Osmotic Energy Conversion Measured in a Single Transmembrane Boron Nitride Nanotube*, Nature **494**, 455 (2013).
[11] A. Siria, M.-L. Bocquet, and L. Bocquet, *New Avenues for the Large-Scale Harvesting of Blue Energy*, Nature Reviews Chemistry **1**, 11 (2017).
[12] S. Hong et al.., *Two-Dimensional Ti3C2Tx MXene Membranes as Nanofluidic Osmotic Power Generators*, ACS Nano **13**, 8917 (2019).
[13] Z. Zhang, S. Yang, P. Zhang, J. Zhang, G. Chen, and X. Feng, *Mechanically Strong MXene/Kevlar Nanofiber Composite Membranes as High-Performance Nanofluidic Osmotic Power Generators*, Nat Commun **10**, 1 (2019).
[14] W. Xin, Z. Zhang, X. Huang, Y. Hu, T. Zhou, C. Zhu, X.-Y. Kong, L. Jiang, and L. Wen, *High-Performance Silk-*

[14] ... *Based Hybrid Membranes Employed for Osmotic Energy Conversion*, Nat Commun **10**, 1 (2019).
[15] Q.-Y. Wu et al., *Salinity-Gradient Power Generation with Ionized Wood Membranes*, Advanced Energy Materials **10**, 1902590 (2020).
[16] Y. Green, Y. Edri, and G. Yossifon, *Asymmetry-Induced Electric Current Rectification in Permselective Systems*, Phys. Rev. E **92**, 033018 (2015).
[17] D. Brogioli, *Extracting Renewable Energy from a Salinity Difference Using a Capacitor*, Phys. Rev. Lett. **103**, 058501 (2009).
[18] A. Meller, L. Nivon, and D. Branton, *Voltage-Driven DNA Translocations through a Nanopore*, Phys. Rev. Lett. **86**, 3435 (2001).
[19] A. Meller, *A New Tool for Cell Signalling Research*, Nat. Nanotechnol. **14**, 732 (2019).
[20] M. Aramesh et al., *Localized Detection of Ions and Biomolecules with a Force-Controlled Scanning Nanopore Microscope*, Nat. Nanotechnol. **14**, 791 (2019).
[21] M. Wanunu, W. Morrison, Y. Rabin, A. Y. Grosberg, and A. Meller, *Electrostatic Focusing of Unlabelled DNA into Nanoscale Pores Using a Salt Gradient*, Nature Nanotechnology **5**, 160 (2010).
[22] M. Wanunu, T. Dadosh, V. Ray, J. Jin, L. McReynolds, and M. Drndić, *Rapid Electronic Detection of Probe-Specific MicroRNAs Using Thin Nanopore Sensors*, Nature Nanotechnology **5**, 807 (2010).
[23] R. Karnik et al., *Electrostatic Control of Ions and Molecules in Nanofluidic Transistors*, Nano Lett. **5**, 943 (2005).
[24] R. Karnik et al., *Rectification of Ionic Current in a Nanofluidic Diode*, Nano Lett. **7**, 547 (2007).
[25] W. Guan, R. Fan, and M. A. Reed, *Field-Effect Reconfigurable Nanofluidic Ionic Diodes*, Nat Commun **2**, 506 (2011).
[26] R. Yan, W. Liang, R. Fan, and P. Yang, *Nanofluidic Diodes Based on Nanotube Heterojunctions*, Nano Lett. **9**, 3820 (2009).
[27] I. Vlassiouk, S. Smirnov, and Z. Siwy, *Nanofluidic Ionic Diodes. Comparison of Analytical and Numerical Solutions*, ACS Nano **2**, 1589 (2008).
[28] I. Vlassiouk, T. R. Kozel, and Z. S. Siwy, *Biosensing with Nanofluidic Diodes*, J. Am. Chem. Soc. **131**, 8211 (2009).
[29] R. A. Lucas and Z. S. Siwy, *Tunable Nanopore Arrays as the Basis for Ionic Circuits*, ACS Appl. Mater. Interfaces **12**, 56622 (2020).
[30] Z. Sarkadi, D. Fertig, Z. Ható, M. Valiskó, and D. Boda, *From Nanotubes to Nanoholes: Scaling of Selectivity in Uniformly Charged Nanopores through the Dukhin Number for 1:1 Electrolytes*, J. Chem. Phys. **154**, 154704 (2021).
[31] D. Fertig, B. Matejczyk, M. Valiskó, D. Gillespie, and D. Boda, *Scaling Behavior of Bipolar Nanopore Rectification with Multivalent Ions*, J. Phys. Chem. C **123**, 28985 (2019).
[32] Y. Qiu, R. A. Lucas, and Z. S. Siwy, *Viscosity and Conductivity Tunable Diode-like Behavior for Meso- and Micropores*, J. Phys. Chem. Lett. **8**, 3846 (2017).
[33] T. Plett, M. L. Thai, J. Cai, I. Vlassiouk, R. M. Penner, and Z. S. Siwy, *Ion Transport in Gel and Gel–Liquid Systems for LiClO4-Doped PMMA at the Meso- and Nanoscales*, Nanoscale **9**, 16232 (2017).
[34] A. Alcaraz et al., *Ion Transport in Confined Geometries below the Nanoscale: Access Resistance Dominates Protein Channel Conductance in Diluted Solutions*, ACS Nano **11**, 10392 (2017).
[35] M. et al., *Scaling Behavior of Ionic Transport in Membrane Nanochannels*, Nano Lett. **18**, 6604 (2018).
[36] S. Faucher et al., *Critical Knowledge Gaps in Mass Transport through Single-Digit Nanopores: A Review and Perspective*, J. Phys. Chem. C **123**, 21309 (2019).
[37] L. Wang et al., *Nanopore-Based Power Generation from Salinity Gradient: Why It Is Not Viable*, ACS Nano **15**, 4093 (2021).
[38] H. B. Park, J. Kamcev, L. M. Robeson, M. Elimelech, and B. D. Freeman, *Maximizing the Right Stuff: The Trade-off between Membrane Permeability and Selectivity*, Science **356**, (2017).
[39] D. Stein, M. Kruithof, and C. Dekker, *Surface-Charge-Governed Ion Transport in Nanofluidic Channels*, Phys. Rev. Lett. **93**, 035901 (2004).
[40] L. Bocquet and E. Charlaix, *Nanofluidics, from Bulk to Interfaces*, Chemical Society Reviews **39**, 1073 (2010).
[41] G. Yossifon and H.-C. Chang, *Changing Nanoslot Ion Flux with a Dynamic Nanocolloid Ion-Selective Filter: Secondary Overlimiting Currents Due to Nanocolloid-Nanoslot Interaction*, Phys. Rev. E **81**, 066317 (2010).
[42] Yoav Green, *Effects of Surface-Charge Regulation, Convection, and Slip Lengths on the Electrical Conductance of Charged Nanopores*, (accepted to Phys. Rev. Fluids).
[43] E. Secchi, A. Niguès, L. Jubin, A. Siria, and L. Bocquet, *Scaling Behavior for Ionic Transport and Its Fluctuations in Individual Carbon Nanotubes*, Phys. Rev. Lett. **116**, 154501 (2016).
[44] P. M. Biesheuvel and M. Z. Bazant, *Analysis of Ionic Conductance of Carbon Nanotubes*, Phys. Rev. E **94**, 050601 (2016).
[45] Y. Uematsu, R. R. Netz, L. Bocquet, and D. J. Bonthuis, *Crossover of the Power-Law Exponent for Carbon Nanotube Conductivity as a Function of Salinity*, J. Phys. Chem. B **122**, 2992 (2018).
[46] Y. Green, *Ion Transport in Nanopores with Highly Overlapping Electric Double Layers*, J. Chem. Phys. **154**, 084705 (2021).
[47] P. B. Peters, R. van Roij, M. Z. Bazant, and P. M. Biesheuvel, *Analysis of Electrolyte Transport through Charged Nanopores*, Phys. Rev. E **93**, 053108 (2016).
[48] R. M. M. Smeets, U. F. Keyser, D. Krapf, M.-Y. Wu, N. H. Dekker, and C. Dekker, *Salt Dependence of Ion Transport and DNA Translocation through Solid-State Nanopores*, Nano Lett. **6**, 89 (2006).
[49] M. Manghi, J. Palmeri, K. Yazda, F. Henn, and V. Jourdain, *Role of Charge Regulation and Flow Slip in the Ionic Conductance of Nanopores: An Analytical Approach*, Phys. Rev. E **98**, 012605 (2018).
[50] O. Schnitzer and E. Yariv, *Electric Conductance of Highly Selective Nanochannels*, Phys. Rev. E **87**, 054301 (2013).
[51] E. Secchi, S. Marbach, A. Niguès, D. Stein, A. Siria, and L. Bocquet, *Massive Radius-Dependent Flow Slippage in Carbon Nanotubes*, Nature **537**, 210 (2016).
[52] Q. Xie, M. A. Alibakhshi, S. Jiao, Z. Xu, M. Hempel, J. Kong, H. G. Park, and C. Duan, *Fast Water Transport in Graphene Nanofluidic Channels*, Nature Nanotech **13**, 238 (2018).
[53] C. Herrero, G. Tocci, S. Merabia, and L. Joly, *Fast Increase of Nanofluidic Slip in Supercooled Water: The Key Role of Dynamics*, Nanoscale **12**, 20396 (2020).
[54] Y. Xie, L. Fu, T. Niehaus, and L. Joly, *Liquid-Solid Slip on Charged Walls: The Dramatic Impact of Charge Distribution*, Phys. Rev. Lett. **125**, 014501 (2020).
[55] Y.-C. Yao, A. Taqieddin, M. A. Alibakhshi, M. Wanunu, N. R. Aluru, and A. Noy, *Strong Electroosmotic Coupling Dominates Ion Conductance of 1.5 Nm Diameter Carbon Nanotube Porins*, ACS Nano **13**, 12851 (2019).
[56] H. J. M. Hijnen, J. van Daalen, and J. A. M. Smit, *The Application of the Space-Charge Model to the Permeability Properties of Charged Microporous Membranes*, Journal of Colloid and Interface Science **107**, 525 (1985).
[57] E. Yariv and J. D. Sherwood, *Application of Schwarz–Christoffel Mapping to the Analysis of Conduction through a Slot*, Proceedings of the Royal Society A: Mathematical, Physical and Engineering Sciences **471**, 20150292 (2015).
[58] Y. Green, R. Eshel, S. Park, and G. Yossifon, *Interplay between Nanochannel and Microchannel Resistances*, Nano Lett. **16**, 2744 (2016).
[59] Y. Green, R. Abu-Rjal, and R. Eshel, *Electrical Resistance of Nanochannel-Microchannel Systems: An Exact Solution*, Phys. Rev. Applied **14**, 014075 (2020).
[60] Y. Noh and N. R. Aluru, *Ion Transport in Electrically Imperfect Nanopores*, ACS Nano **14**, 10518 (2020).
[61] Y. Green, *Conditions for Electroneutrality Breakdown in Nanopores*, J. Chem. Phys. **155**, 184701 (2021).